\documentclass[sigconf]{acmart}
\AtBeginDocument{%
  }

\usepackage{framed}

\usepackage[ruled, linesnumbered]{algorithm2e}

\SetKwComment{Comment}{/* }{ */}
\SetKwRepeat{Do}{do}{while}

\usepackage{xcolor}

\newcommand{\green}[1]{\textcolor{black}{#1}}

\usepackage{makecell}

\usepackage{bbding}

\usepackage{multirow}

\usepackage{enumitem}

\usepackage{pifont}

\usepackage[title]{appendix}

\usepackage{tcolorbox}
\tcbuselibrary{breakable}

\newcommand{\mname}{AgentSentinel\xspace}
\newcommand{\bname}{BadComputerUse\xspace}

\begin{document}

\title{AgentSentinel: An End-to-End and Real-Time Security Defense Framework for Computer-Use Agents}

\author{Haitao Hu}
\affiliation{%
  \institution{ShanghaiTech University}
  \city{Shanghai}
  \country{China}}
\email{huht2022@shanghaitech.edu.cn}

\author{Peng Chen}
\affiliation{%
  \institution{Independent Researcher}
  \city{Shanghai}
  \country{China}}
\email{spinpx@gmail.com}

\author{Yanpeng Zhao}
\affiliation{%
  \institution{Independent Researcher}
  \city{Beijing}
  \country{China}}
\email{yannzhao.ed@gmail.com}

\author{Yuqi Chen}
\authornote{Corresponding author}
\affiliation{%
  \institution{ShanghaiTech University}
  \city{Shanghai}
  \country{China}}
\email{chenyq@shanghaitech.edu.cn}


\begin{abstract}
Large Language Models (LLMs) have been increasingly integrated into computer-use agents, which can autonomously operate tools on a user's computer to accomplish complex tasks. However, due to the inherently unstable and unpredictable nature of LLM outputs, they may issue unintended tool commands or incorrect inputs, leading to potentially harmful operations. Unlike traditional security risks stemming from insecure user prompts, tool execution results from LLM-driven decisions introduce new and unique security challenges. These vulnerabilities span across all components of a computer-use agent. To mitigate these risks, we propose \mname, an end-to-end, real-time defense framework designed to mitigate potential security threats on a user's computer. \mname intercepts all sensitive operations within agent-related services and halts execution until a comprehensive security audit is completed. Our security auditing mechanism introduces a novel inspection process that correlates the current task context with system traces generated during task execution. To thoroughly evaluate \mname, we present BadComputerUse, a benchmark consisting of 60 diverse attack scenarios across six attack categories. The benchmark demonstrates a 87\% average attack success rate on four state-of-the-art LLMs. Our evaluation shows that \mname achieves an average defense success rate of 79.6\%, significantly outperforming all baseline defenses.
\end{abstract}


\maketitle

\section{Introduction}

Large Language Models (LLMs), such as GPT-4o~\cite{gpt-4o} and Claude 3.7 Sonnet~\cite{claude37}, represent the most advanced natural language processing technologies to date. These models accept user prompts, describing problems, requirements, or tasks, and generate reasonable and contextually appropriate responses. Due to their powerful task-processing capabilities, LLMs have been widely adopted across numerous domains and integrated into a variety of LLM-based applications. To further enhance their capabilities, tools are often integrated with LLMs, enabling them to interact with the external world—these enhanced systems are commonly referred to as LLM-based agents. Additionally, the development of the Model Context Protocol (MCP)~\cite{mcp} has significantly accelerated the integration of LLMs into such agents.

Among the diverse range of agents, one of particular significance has emerged: the computer-use agent. These agents allow LLMs to take full control of a user's computer environment and perform complex tasks much like a human user. For instance, Anthropic first introduced computer-use capabilities in Claude 3.5 Sonnet~\cite{claude35}. OpenAI has since followed with its Operator framework and the Computer-Using Agent model~\cite{operator}. Similarly, Manus~\cite{manus} is a general-purpose AI agent that can interact with a cloud-based computer environment. As a result, the ability to perform real-world computer interactions—referred to as computer-use has become a critical capability for next-generation LLM-based agents.

The use of computer-use agents inevitably introduces new vulnerabilities and significantly expands the attack surface due to their ability to interact with the computer environment. Specifically, in scenarios without an adversary, the unpredictable nature of LLMs can result in error-prone or overly complex behavior. If an LLM gains control of a user's computer and executes any operation it deems appropriate, this raises immediate and serious concerns. As a result, incorrect tool use may compromise the integrity of the user's operating system and, in severe cases, cause irreversible damage. Therefore, it is critical to intercept such risky actions promptly to prevent further harm. 

In adversarial settings, both the agent infrastructure and the underlying operating system are susceptible to attacks targeting different components of the computer-use agent. To compromise the agent infrastructure, an attacker could issue seemingly benign instructions designed to exfiltrate private information or tamper with critical components. For example, instructing the agent to read configuration files that define its behaviors \cite{manus-leak} or to terminate a process that supports the agent's core functionality. Alarmingly, most LLMs are willing to comply with such instructions without hesitation. To compromise the operating system, a third-party adversary could exploit the agent’s interaction with external tools. For instance, if the agent fetches data using a tool whose source is controlled by an attacker, the attacker can inject malicious payloads into the tool results. These malicious results then infect the task context maintained by the agent, potentially leading to command injection or other harmful behaviors.

While numerous techniques have been proposed to mitigate the risks associated with tool use, these approaches are unfortunately ineffective when applied to computer-use agents. For example, Zhang et al.~\cite{zhang2025agent} propose instruction-level defenses aimed at mitigating incorrect outputs generated by LLMs. However, these methods often lack robustness against the sophisticated threats posed by malicious attackers in the context of computer-use agents. Another line of work (e.g.,~\cite{wu2024isolategpt}) explores sandboxing and isolation strategies. Unfortunately, computer-use agents typically possess broad access to and control over most tools on a user's system, violating the core assumptions underlying these techniques. Finally, security-enhanced agent architectures~\cite{kim2025prompt} are also inadequate for computer-use agents. Re-architecting or hardening individual components of these agents is impractical, given the diversity and rapid proliferation of such agents in real-world deployments. 

To address the aforementioned issues, we propose \mname, an end-to-end, real-time security defense framework for computer-use agent systems. \mname passively traces each tool invocation by dynamically collecting relevant system events. When a sensitive operation is detected, it is suspended, and detailed information is presented to a security auditor, who then determines the appropriate enforcement actions. It is worth noting that our approach addresses all threats at the endpoint, regardless of the specific vulnerabilities exploited within the computer-use agent. This design minimizes interference from tool behavior on the agent’s execution.

\mname comprises four core components. The first two components, namely the agent instrumentation component and the monitoring server component, form a novel client-server architecture. In this architecture, the monitoring server protects the agent operating as a client. The agent instrumentation component provides a minimally intrusive and composable security module that can be effortlessly integrated into existing agent systems. It performs three primary functions: (1) establishing a trustworthy communication channel and transmitting basic agent information to a monitoring server; (2) collecting dynamic task context and sending it to the monitor; and (3) delivering security alerts back to the agent when a threat is identified. 

In addition, we use a tracer component to trace and enforce the sensitive operations initiated by the agent. Specifically, upon receiving a connection request from the agent client, the monitoring server configures and activates the tracer component, initially operating in a passive monitoring mode. The tracer component silently logs all sensitive actions, and immediately halts any process against our predefined enforcement policy. Once a process is suspended, a security audit module analyzes the corresponding operation, taking into account the task context and associated event traces. If the operation is deemed safe, the process is resumed; otherwise, it is terminated to prevent further harm. The agent client is notified via a security alert if a tool failure is detected.

Finally, we propose a novel \textit{task-context-aware security audit component} designed to detect potential risks within the computer-use agent. The security auditor comprises two complementary components: a \textit{rule-based auditor} that enforces predefined security policies, and an \textit{LLM-based auditor} that leverages the reasoning capabilities of large language models to assess complex scenarios. To enhance the efficiency and responsiveness of the LLM-based auditor, we propose a \textit{Query-Per-Second (QPS) optimizer} that regulates audit latency and ensures minimal disruption to normal task execution. Furthermore, we implement a \textit{security query cache} that classifies previously verified operations into three hierarchical categories: \textit{universally safe}, \textit{task-specific safe}, and \textit{safe-for-next-time}. This caching mechanism effectively reduces redundant queries and auditing overhead by pruning unnecessary event traces and reusing prior verification results.

Our contributions are summarized as follows.
\begin{itemize}[topsep=2pt]
    \item We propose \mname \footnote{AgentSentinel available at \url{https://github.com/m4p1e/agent-sentinel}}, a novel end-to-end, real-time security defense framework for computer-use agents. To the best of our knowledge, this is the first, end-to-end and real-time defense framework for computer-use agents.
    \item  We present BadComputerUse, a benchmark consisting of 60 diverse attack scenarios across seven attack types, which demonstrates an average attack success rate of 87\% on four state-of-the-art LLMs. 
    \item Our evaluation results show that \mname achieves an average defense success rate of 79.6\%, significantly outperforming all baseline defenses.
\end{itemize}

\section{Background \& Problem Statement}

The computer-use agent is a new class of LLM-based agents. It interacts directly with a user's operating system (OS). This ability allows the agent to use a wide range of existing tools on the OS. Additionally, it can access various interfaces connected to peripherals, such as the monitor, keyboard, and mouse. This grants the agent extensive task execution possibilities. For example, a traditional LLM-based agent might use an external web search plugin to get information. In contrast, a computer-use agent can launch a browser, navigate to a web page, and type URLs into the address bar. However, Xia et al.~\cite{xie2024osworld} show that current computer-use capabilities of LLMs are still insufficient for fully automating complex computer tasks.

\begin{figure}[t]
    \centering
    \includegraphics[width=1\linewidth]{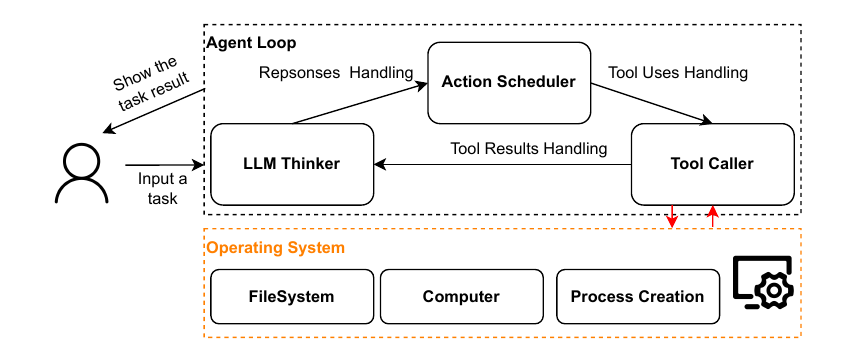}
    \caption{General architecture of a computer-use agent.}
    \label{fig:agent-arch}
\end{figure}

Figure \ref{fig:agent-arch} presents an illustrative architecture of a computer-use agent, comprising three capabilities of tool calling: filesystem access, computer control, and process creation. The agent processes user task inputs within a tightly coupled feedback loop while maintaining a dynamic task context. This agent loop involves three key components: 1) a LLM thinker module, where the LLM responds to the task context by generating a sequence of actions representing the next execution steps; 2) an action scheduler module, which updates the task context with the LLM’s response and schedules tool invocations accordingly; and 3) a tool caller module, which invokes the requested tools and updates the context with the resulting outputs. The loop continues iteratively until no new tool use requests are generated in the LLM’s response.

\subsection{Computer-Use Agents}\label{bg:agent-intro}

\subsection{Existing Attack Surfaces in LLMs}

\textbf{Prompt Injection.} 
\label{sec:prompt_injection}
The outputs of LLMs are mainly controlled by input prompts, which can be broadly categorized into developer instructions and user prompts. Due to the inherent instruction-following nature of LLMs, the boundary between these categories is often ambiguous. Adversaries can exploit this by creating malicious prompts. These malicious prompts can override developer-imposed constraints, leading to unauthorized or harmful actions. Based on attack objectives, prompt injection techniques are used in $\mathnormal{J}\!\emph{ailbreaking}$ attacks~\cite{kumar2023certifying,zou2023universal,jia2024improved,wang2024attngcg} and \emph{privacy leakage} attacks~\cite{hui2024pleak,edemacu2024privacy}. $\mathnormal{J}\!\emph{ailbreaking}$ attacks aim to circumvent safety alignments or context restrictions, such as chat history, to elicit harmful or unintended outputs. Privacy leakage attacks focus on extracting sensitive information, such as personally identifiable information (PII) or hidden system prompts.

\noindent\textbf{Backdoor Attacks.} 
Third-party LLMs are untrusted because of the presence of backdoors~\cite{zhou2025survey}. These backdoors can be implanted during model training~\cite{wan2023poisoning} or through system prompt manipulation~\cite{zhang2024instruction}. An adversary can activate a backdoor by including a specific trigger within an otherwise benign prompt. When the LLM detects this trigger, it deviates from its expected behavior. As a result, the LLM produces malicious or unexpected outputs.

\noindent\textbf{Hallucination Risks.} Prior studies ~\cite{zhang2023siren,huang2025survey} show that LLMs often generate hallucinations: outputs that are fluent but factually incorrect. These hallucinations usually arise from ambiguous queries or limitations in training data. When these fabricated outputs are treated as factual, they pose serious security risks, especially in critical or sensitive applications.

\subsection{New Attack Surfaces in Computer-Use Agents}

\textbf{Indirect Prompt Injection.}
When an LLM has the capability to retrieve external resources via integrated tools, the retrieved content becomes susceptible to \emph{indirect prompt injection} attacks~\cite{greshake2023not}. Unlike the direct prompt injection methods examined in~\autoref{sec:prompt_injection}, this vulnerability emerges when the LLM automatically incorporates unverified external data into its processing pipeline. If these external sources are not rigorously validated, malicious inputs can trigger prompt injection vulnerabilities through seemingly benign channels.

\noindent\textbf{OS-Targeted Backdoor Attacks.}
In a purely conversational setting, the outputs of a backdoor may have limited real-world impact. However, when such outputs are consumed by an agent system, they may influence the underlying operating system, especially when the backdoor consistently generates dangerous tool commands~\cite{wang2024badagent}. This type of OS-targeted backdoor can elevate a conventional LLM backdoor into a full system compromise. Consequently, it poses critical security threats to devices or infrastructures that are running LLM Agents.

\noindent\textbf{Hallucination Exploitation.}
Hallucinations can serve as potent attack vectors in agent-based systems. For instance, if an LLM hallucinates a non-existent package while responding to a Python package recommendation task, an attacker could upload a malicious package with that hallucinated name to a public repository. This could lead to the unintended installation of the malicious package by agents executing software installation tasks, thereby introducing serious security vulnerabilities.

\noindent\textbf{Untrustworthy Agent Execution Environment.} An untrustworthy agent execution environment may expose agents to malicious local resources. Agents often depend on external tools, which could include locally installed software or dynamically downloaded dependencies. If a required tool is missing, the agent might retrieve it from an unverified third-party source. Both pre-existing and newly installed dependencies may contain malicious components. Additionally, some third-party agent systems~\cite{autogpt, lu2024omniparserpurevisionbased,operator} utilize isolated execution environments that bundle all necessary dependencies (e.g., customized browsers or virtual machines). Since these tools operate internally within the agent’s execution environment, users have no reliable way to verify their trustworthiness.

\begin{figure*}[t]
    \centering
    \includegraphics[scale=0.5]{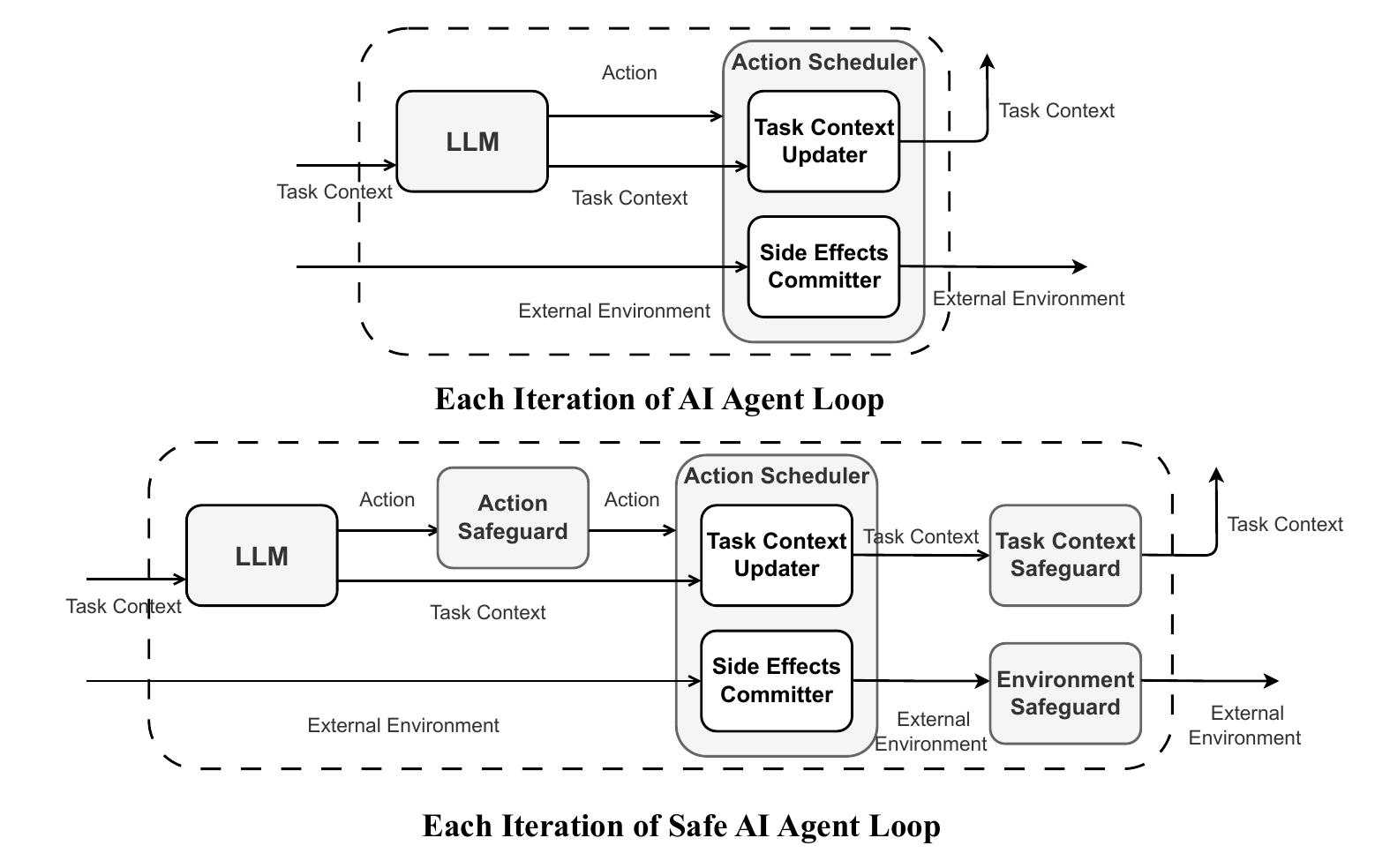}
    \caption{\green{Composable state transformer architecture for AI agent loop.}}
    \label{fig:problem-st}
\end{figure*}

\subsection{Challenges for Securing Computer-Use Agents}
To effectively defend against the aforementioned attacks in real time, several critical challenges must be addressed to secure the use of computer-use agents.

\noindent\green{\textbf{Challenge 1: Traditional prompt and task context guards fail to comprehensively detect threats.}
An agent's task context includes elements such as chat history and tool use records. However, tool use records typically include only the tool’s name, inputs, and outputs, without detailed execution traces within the tools themselves. Consequently, context-based audits often fail to detect sensitive operations based solely on these limited contextual cues. Furthermore, a single tool invocation may trigger multiple underlying processes at the operating system level, whose sensitive behaviors remain hidden from such audit mechanisms. Therefore, it is essential to capture sensitive operations occurring during tool use through more appropriate and effective means.}

\noindent\green{\textbf{Our Solution:} \mname analyzes both internal task context and external side effects generated by the agent, enabling more thorough and context-aware threat detection.}

\noindent\green{\textbf{Challenge 2: Side effects from computer-use agents can affect multiple components of external environments.} Computer-use agents may invoke arbitrary OS utilities, enabling them to leverage any existing tool within the operating system rather than being restricted to tools specified in system prompts. Consequently, constraining tool use through limited security policies is impractical. Furthermore, generating specific safety rules for each individual tool is also unfeasible. A more general enforcement strategy is needed to address all potential tool-use scenarios.}

\noindent\green{\textbf{Our Solution:} \mname utilizes a system tracer to capture operations relevant to tool use, including process control, file operations, and network access. Based on these collected operations, \mname models potential security threats associated with tool execution.}

\noindent\green{\textbf{Challenge 3: Real-time threat defense for agents must be efficient to avoid excessive latency.}
A real-time defense approach must detect sensitive behaviors and intervene before their execution. Upon detection, it halts the behavior, conducts a security audit, and either resumes or terminates the process based on the audit results. This defense process introduces execution delays due to auditing and enforcement (i.e., stopping and resuming processes). However, computer-use agent often impose strict timeouts on tool use to maintain responsiveness. A critical challenge arises when the audit latency exceeds these timeouts, potentially disrupting normal agent operations—especially if a single tool use triggers numerous sensitive operations. Consequently, it is crucial to design mechanisms that either accelerate the security audit process or provide flexible control over its execution time.}

\noindent\green{\textbf{Our Solution:} \mname integrates rule-based and LLM-based security audit approaches. To further reduce the latency of LLM-based audits, we introduce a novel security query cache to accelerate LLM-based audits, along with a queries-per-second (QPS) optimizer that enforces strict time constraints for each audit decision.}

\begin{figure*}[t]
\begin{center}
    \includegraphics[scale=0.6]{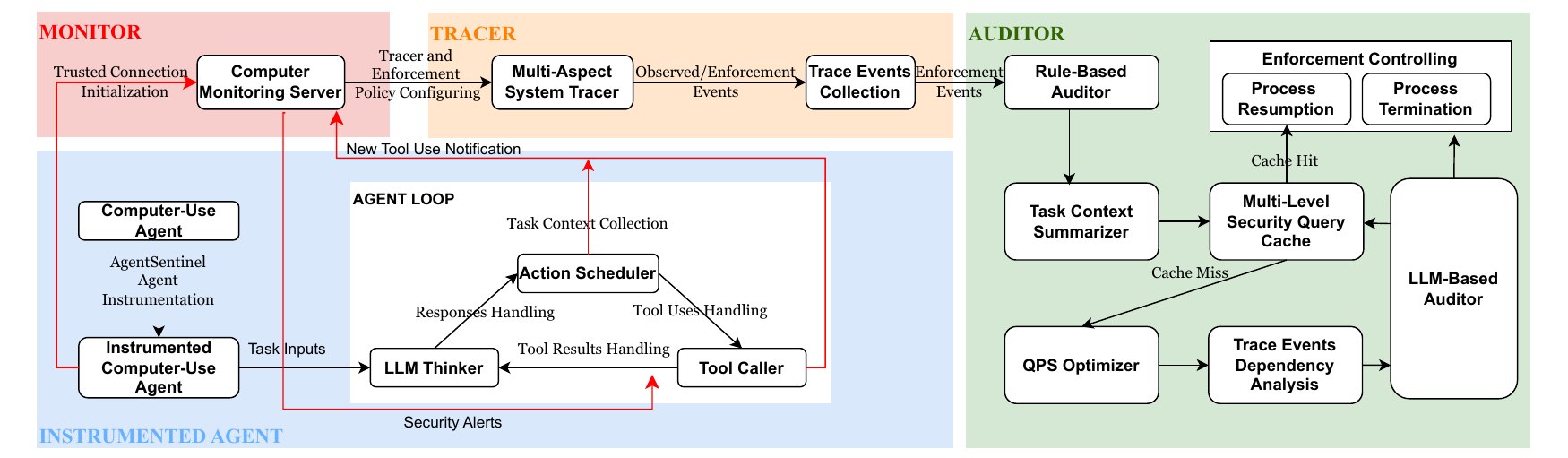}
\end{center}
\caption{Overview of \mname.}
\label{fig:sandbox-arch}
\end{figure*}

\section{Problem Statement}\label{sec-prob-st}

\green{\textbf{Problem Formulation.}
\cite{he2025security} models LLMs and AI agents as composable state transformers. Extending this formalism, we introduce security mechanisms into the state transformer architecture, as depicted in Figure \ref{fig:problem-st}. Formally,
let $\mathcal{C}$ denote the internal state space representing the agent's task context, and let $\mathcal{A}$ denote the action state space representing the agent's possible actions. We can define a state transformer of LLM as a function $\mathcal{T}_\mathrm{LLM} \colon~\mathcal{C} \to \mathcal{A}$,  which represents how the LLM processes a given task context and outputs the next action. Furthermore, let $\mathcal{O}$ denote the external state space representing the environment in which the agent operates. We define the state transformer of the agent's action scheduler as two functions: $\mathcal{T}_\mathrm{int} \colon \mathcal{A} \times \mathcal{C} \to \mathcal{C}$ and $\mathcal{T}_\mathrm{ext} \colon \mathcal{A} \times \mathcal{O} \to \mathcal{O}$. The function $\mathcal{T}_\mathrm{int}$ denotes how the agent performs actions and updates the task context, while the function $\mathcal{T}_\mathrm{ext}$ denotes how actions affect the external environment. Finally, we define the state transformer of each iteration of the agent loop as a composite function $\mathcal{T}\colon~\mathcal{C} \times \mathcal{O} \to \mathcal{C} \times \mathcal{O}$ 
\begin{equation}
    \label{common-trans-func}
    \mathcal{T} (c, o) = \left[\mathcal{T}_\mathrm{int}(a, c),~~ \mathcal{T}_\mathrm{ext}(a, o)\right]
\end{equation}
where $a = \mathcal{T}_\mathrm{LLM}(c)$.}

\green{We also consider the security threat defense mechanism for computer-use agents as a new composable component within the agent loop, consisting of three functions:
\begin{enumerate}
    \item A function $\mathcal{T}_\mathrm{secAct}: \mathcal{A} \to \mathcal{A}$ for securing action execution.
    \item A function $\mathcal{T}_\mathrm{secCtx}: \mathcal{A} \times \mathcal{C} \to \mathcal{C}$ for securing task context updates.
    \item A function $\mathcal{T}_\mathrm{secEff}: \mathcal{A} \times \mathcal{O} \to \mathcal{O}$ for securing side effect commits.
\end{enumerate}
We can then define the safe state transformer of each iteration of the agent loop as a new composite function:
\begin{equation}
    \label{safe-common-trans-func}
    \mathcal{T}_\mathrm{safe} (c, o)= \left[
    \begin{aligned}
        &\mathcal{T}_\mathrm{secCtx}(a', \mathcal{T}_\mathrm{int}(a', c)), \\
        &\mathcal{T}_\mathrm{secEff}(a', \mathcal{T}_\mathrm{ext}(a', o))
    \end{aligned}\right]
\end{equation}
where $a' = \mathcal{T}_\mathrm{secAct}(\mathcal{T}_\mathrm{LLM}(c))$.
Consequently, a security defense mechanism must provide concrete definitions for the three securing functions in Equation~\ref{safe-common-trans-func}.}

\noindent\green{\textbf{Our Solutions.} In this paper, we define a composable security threat defense component as $(\mathcal{T}_\mathrm{secAct}', ~\mathcal{T}_\mathrm{secCtx}',~ \mathcal{T}_\mathrm{secEff}')$ where 
\begin{enumerate}
    \item For any action $a \in \mathcal{A}$, we have $\mathcal{T}_\mathrm{secAct}'(a) = a$. Our defense mechanism does not directly interfere with agent's execution flow to avoid affecting agent capability. In other words, the agent remains unrestricted in issuing tool calls.
    \item For any action $a \in \mathcal{A}$ and task context $c \in \mathcal{C}$, we have
    \[
        \mathcal{T}_\mathrm{secCtx}'(a, c) = \left\{
        \begin{aligned}
            &\mathrm{appendSecurityAlert}(c) && \textrm{if}~\neg \mathrm{isSafeAction}'(a)\\ 
            &c && \textrm{otherwise}.
        \end{aligned}\right.
    \]
    where the operation \emph{appendSecurityAlert} appends a security alert to the task context, and the operator \emph{isSafeAction} is defined as:
    \[
        \mathrm{isSafeAction}'(a) =  \forall e ~ \mathrm{isSafeOp}(e)
    \]
    where $e$ represents each sensitive operation produced within the action $a$, and the operator \emph{isSafeOp} determines whether the operation is safe. Our defense mechanism appends security alerts to the task context only when security threats are detected during action execution.
    \item For any action $a \in \mathcal{A}$ and external environment $o \in \mathcal{O}$, we have
    \[
        \mathcal{T}_\mathrm{secEff}'(a, o) = \left\{
        \begin{aligned}
            &\mathrm{PE}(o) && \textrm{if}~\neg \mathrm{isSafeAction}'(a)\\ 
            &o && \textrm{otherwise}.
        \end{aligned}\right.
    \]
    where \emph{PE} commits the partial execution effects (generated prior to tool termination) to the external environment. Our defense mechanism immediately terminates action execution when security threats are detected. 
\end{enumerate}}

\noindent\textbf{Threat Model.} We assume that an attacker can access the computer-use agent system through any available interface. The potential attacks are categorized into six classes:

\begin{enumerate}
    \item  \textbf{Direct Prompt Injection}: An attacker appends malicious instructions to a legitimate user task prompt. 
    \item  \textbf{Agent Infrastructure Attack}: A user submits tasks designed to compromise the agent infrastructure. 
    \item  \textbf{Backdoor Attack}: When a third-party LLM contains a hidden backdoor, an attacker may exploit it by triggering malicious behavior through specific user inputs, potentially enabling harmful operations on the underlying operating system.
    \item  \textbf{Bad Tool Result Attack}: If the agent retrieves attacker-controlled external resources during tool use, these resources may deliver malicious or manipulative data that cause prompt injection into the task context. 
    \item \textbf{Hallucination Exploitation}: A benign user may input tasks that are prone to hallucinations, potentially leading to unintended agent behavior. 
    \item  \textbf{Malicious Agent Execution Environment:} When using third-party integrated agent environments containing malicious applications, these payloads may execute during task handling. Attackers may also embed such malicious applications within popular tools to increase their usage likelihood.
\end{enumerate}

\section{Design}

In this section, we present \textbf{\mname}, a novel framework designed to detect real-time security threats in computer-use agents. Figure \ref{fig:sandbox-arch} presents an overview of \mname's architecture, which comprises four core components designed to address key challenges in securing computer-use agents. 

\begin{enumerate}
    \item \green{\textbf{Context-Aware Agent Monitoring:} \mname instruments the computer-use agent to initialize communication channel and dynamically collect task context (blue components). Both agent basic information and task context are transmitted via an agent-to-monitor protocol to the monitoring server (red components).}
    \item \green{\textbf{System-Level Tracing:} \mname utilizes a multi-aspect system tracer to capture and suspend sensitive operations (orange components).}
    \item \green{\textbf{Adaptive Hybrid Threat Auditing:} \mname integrates a modular security audit approach designed to efficiently identify real-time threats associated with tool execution (green components).}
\end{enumerate}

\SetAlFnt{\small}
\begin{algorithm}[t]
    \caption{Instrumented Agent}\label{alg:inst-agent-prog}
    {\color{red}$C \gets \mathrm{New\mname Client}(\mathit{AgentBasicInformation})$\;}
    $I \gets \mathrm{GetTaskInput}()$\;
    {\color{red}$C.\mathrm{AddUserMessage}(I)$\;}
    $M \gets (I)$\;
    $T \gets \mathrm{GetAvailableTools}()$\;
    \Do{$M \neq M'$}{
      $A \gets \mathrm{GetActionsFromLLM}(M, T)$\;
      $M' = M$\;  
      \For {$a \in A$}{
        \eIf{$\mathrm{IsToolUse}(a)$}{
            {\color{red}$C.\mathrm{NotifyNewToolUse}(a)$\;}
            $r = \mathrm{RunTool}(a)$\;
            {\color{red}$C.\mathrm{AddUserMessage}(r)$\;}
            {\color{red}\If{$\mathrm{IsToolError}(r)$}{
                $r = C.\mathrm{TryToAppendSecurityAlert(r)}$\;}}
            $M = M + r$\;
            
        }{
            {\color{red}$C.\mathrm{AddAgentMessage}(a)$\;}
        }
      }
    }
\end{algorithm}

\subsection{Computer-Use Agent Instrumentation} \label{design-1}

To obtain the task context and determine the next action from a running agent, we first instrument the original agent program to be capable of communicating with the monitoring server.  The general structure of the instrumented agent is presented in Algorithm \ref{alg:inst-agent-prog}, where the red-highlighted lines indicate modifications to the original code. At line 1, the agent initializes a \mname client by establishing a connection with the monitoring server, which enables the server to retrieve essential information such as the agent’s process ID and its dependent files. The initial task input is collected at line 3. For each tool-use action, the client notifies the server of an impending tool invocation. All task context messages—gathered at lines 3, 12, and 19—are transmitted at line 11, but only when a tool-use action is imminent. This notification process is non-blocking, allowing the \mname server to audit the tool invocation passively and mitigate potential security threats in real time. If the auditor enforces a restriction on the tool use, the corresponding security alerts are appended to the tool’s error results. This is the only instance in which the \mname client modifies the task context. Overall, the \mname client introduces minimal intrusion to the original agent architecture, enabling users to integrate it with minimal code refactoring. These helper functions can also be embedded into software development kits (SDKs) provided by popular LLM platforms.

\subsection{Agent-to-Monitor Communication Protocol}\label{design-2}

We design a robust, attack-resistant communication protocol to establish a trustworthy channel between the agent client and the monitoring server. We refer to this as the \textbf{Agent-to-Monitor (A2M) Communication Protocol}. The protocol is detailed in Algorithm~\ref{alg:a2m-proto}, where the communication procedures for both parties are specified independently. Both sides are required to execute their respective procedures without error; any failure will result in a panic response, thereby halting further communication. This strict failure model ensures the protocol operates deterministically and securely. Moreover, the sequence of procedure calls is rigidly defined: 1) The monitor initializes a listening server and waits for incoming connection requests from agent clients. 2) During initialization, the agent attempts to connect to the monitor using a provided server descriptor (e.g., URL). 3) A reliable channel is established via a $\mathit{SayHello}$ handshake protocol. 4) All subsequent data communication is initiated by the agent and concluded with the monitor's response.

Communication is driven by client requests, each consisting of an \emph{operand} and associated \emph{data}. The A2M protocol currently supports four operand types: $\mathit{op}_\mathrm{connect}$, $\mathit{op}_\mathrm{startPassiveTracing}$, $\mathit{op}_\mathrm{sendNewToolUse}$ and $\mathit{op}_\mathrm{getEnforcementInfo}$. The monitor server handles operands in a stateful manner. The initial request must be $\mathit{op}_\mathrm{connect}$, which contains identifying information about the agent. The $\mathit{op}_\mathrm{startPassiveTracing}$ request, which enables system-level tracing and auditing, is permitted only after the initial connection is established. The remaining two request types are valid only once tracing is active. Importantly, the protocol does not permit any operand that disables tracing once it has been enabled. This design defends against malicious attempts to terminate security monitoring, which would compromise the trust model after task inputs have been provided. By enforcing this unidirectional trust boundary, \mname ensures continuous protection even in adversarial environments.

\begin{algorithm}[t]
    \caption{A2M Protocol}\label{alg:a2m-proto}
    \SetKwProg{myproc}{Procedure}{}{}
    
    \myproc{$\mathtt{AgentClient}(\mathit{serverDesc}, data_{\mathrm{AgentBasicInfo}})$}{
        {$C \gets \mathrm{ConnectAndSayHello}(\mathit{serverDesc})$}\;
        {$\mathrm{SendClientRequest(C, op_{\mathrm{connect}}, data_{\mathrm{AgentBasicInfo}})}$}\;
        \While{$\mathit{true}$}{
            $op, data \gets \mathrm{GenerateNewClientRequest}()$\;
            {$req \gets \mathrm{SendClientRequest(C, op, data)}$}\;
            {$\mathrm{ConsumeServerResponse}(op, req)$}\;
        }
    }

    \myproc{$\mathtt{MonitoringServer}()$}{
        $S \gets \mathrm{InitListeningServerFromServerDescriptor}()$\;
        $C \gets \mathrm{WaitClientConnectAndSayHello}(S)$\;        
        $\mathit{status} \gets 0$\;
        
        \While{$\mathit{true}$}{
            $op, data \gets \mathrm{WaitNewClientRequest}()$\;
            \If{$\mathit{status} = 0$ and $\mathit{op} = \mathit{op}_\mathrm{connected}$}{
                $\mathit{status} \gets 1$\;
            }

            \If{$\mathit{status} = 1$ and $\mathit{op} = \mathit{op}_\mathrm{startPassiveTracing}$}{
                $\mathit{status} \gets 2$\;
            }

            $\mathrm{HandleClientRequest}(\mathit{status}, C, op, data)$\;
        }
    }
\end{algorithm}

\begin{figure*}[t]
\begin{center}
    \includegraphics[scale=0.61]{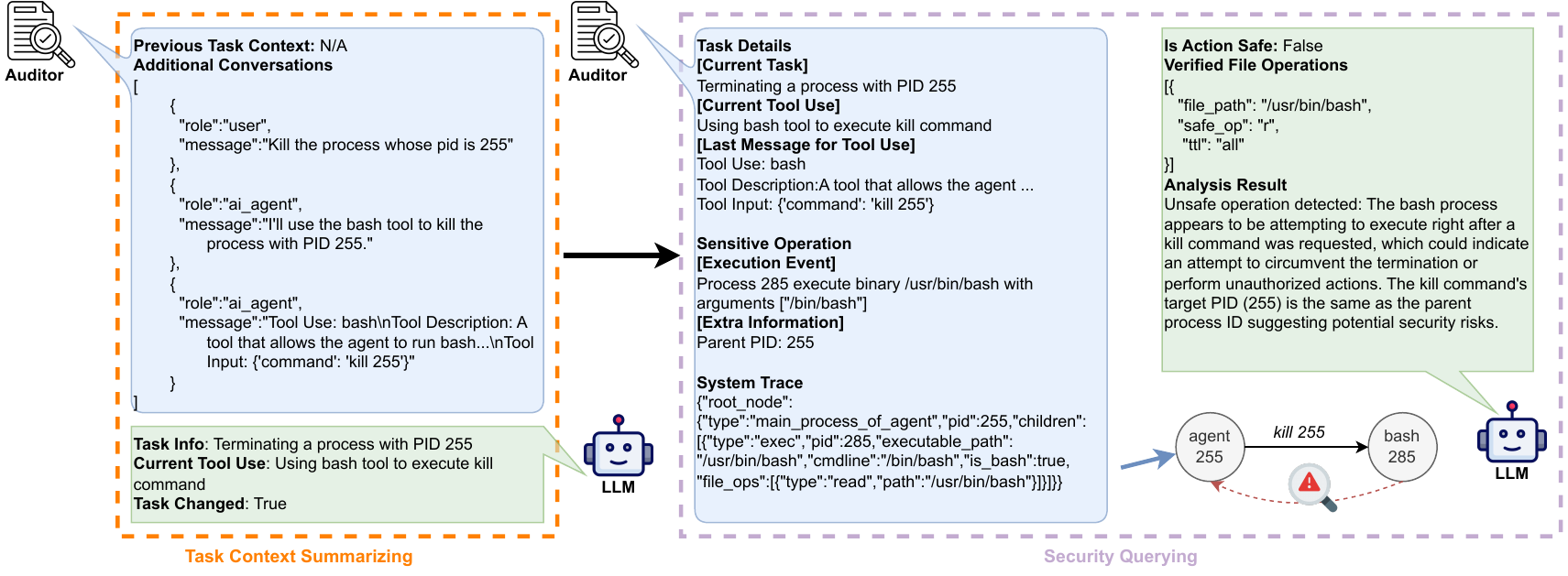}
\end{center}
\caption{Task context summarizing and security querying of the LLM-based auditor.}
\label{fig:audit}
\end{figure*}

\subsection{System Tracer and Enforcer} \label{design-3}

To identify side effects introduced by tool use in computer-use agents, we instrument 16 trace probes within critical components of the underlying operating system. These probes cover process control, file operations, and network access. Each trace probe passively logs observed events to an event collector. If an observed event violates the enforcement policy associated with the probe, the responsible process is subject to enforcement actions such as suspension or termination. We refer to such an observed event as an enforcement event. The enforcement policies on all trace probes are configurable, enabling selective enforcement for specific events.

\noindent\textbf{Process Control Probes.} We focus on four key process-related operations: \emph{fork}, \emph{exec}, \emph{kill}, and \emph{exit}. Tool use frequently involves spawning multiple processes. For instance, when an agent uses the command cat \texttt{/tmp/server.log} via a shell, it invokes both the bash and cat processes. Specifically, the agent forks a process to execute bash, which in turn interprets the command and executes cat. The fork and exec probes allow us to identify the precise process tree and binary dependencies involved in a tool use. Meanwhile, kill and exit probes capture signal exchanges and program termination statuses, helping infer the overall program state. To manage scope and reduce noise, process control probes maintain two sets: interesting processes and enforced processes. A process is marked as interesting if it is the main agent process or one of its descendants. Enforcement is governed by a tunable threshold—maximum enforced process level—where the main agent starts at level 1, its children at level 2, and so forth. A process is subject to enforcement if its level is less than or equal to this threshold.

\noindent\textbf{Filesystem Access Probes.} We instrument three key file operations: \emph{open}, \emph{remove}, and \emph{rename}. Tool use often involves more file accesses than one might expect, including both explicit (e.g., reading \texttt{/tmp/server.log}) and implicit (e.g., loading binaries like bash and cat, or reading configuration files). File-level operations can generate a large volume of low-level events; for example, writing a large file through buffered I/O may trigger multiple write events. To reduce overhead, we focus on tracing file open events, which provide sufficient insight via access modes (e.g., read/write). Additionally, remove and rename probes help track structural changes in the filesystem, excluding content modifications.

\noindent\textbf{Network Access Probes.} We focus on internet-related activities (e.g., IPv4/IPv6) and three core operations: \emph{connect}, \emph{listen}, and \emph{accept}. A network-enabled agent may both retrieve data and expose services. By tracing connect operations, we capture outbound communication capabilities and destination addresses. Conversely, listen and accept traces reveal server-side behavior and inbound connections from remote clients. Furthermore, we trace DNS resolution events to maintain a DNS mapping table. This allows reverse-mapping IP addresses to their original domains. For example, resolving an HTTP request involves first querying a domain via DNS, then initiating a TCP connection. With the DNS table, we can contextualize remote addresses using their associated domains, aiding in network source analysis.

\subsection{Real-time Security Risks Auditing}\label{sec:audit} To identify and mitigate real-time security threats triggered by the agent during tool use, we propose an automated, modular and configurable security auditor that blocks risky operations upon threat detection. \green{Importantly, this auditor is driven by sensitive trace events, instead of directly interfering with every function call from the agent.} The security auditing procedure is detailed in Algorithm~\ref{alg:audit}. This auditor accepts four arguments: the enforcement event $e$, the event trace $\mathit{tr}$ collected during the current tool session, new chat messages $\mathit{msgs}$ from the agent client, and a boolean flag $\mathit{hasNewToolUse}$ indicating whether a new tool is being used by the agent. If the enforcement event is deemed safe, the auditor resumes the corresponding process; otherwise, it terminates the process. The auditing workflow consists of four stages: 1) The rule-based auditor applies predefined policies to evaluate enforcement events, leveraging efficient pattern matching techniques for policy enforcement. 2) When a new tool use is detected, chat messages are incrementally summarized to construct the task context. 3) If the task has changed, verified task-dependent operations are flushed from the cache. The auditor checks for a cache hit and resumes the process immediately if one is found. 4) Constructs a security query using the enforcement event, its dependent event trace, and the summarized task context. This query is evaluated by an LLM-based auditor to return a decision (resumption or termination), the set of verified operations, and an explanation. Figure~\ref{fig:audit} illustrates a task where the agent attempts to kill a specific process. The chat messages are synthesized into a concise task summary, and the security query—comprising the task summary, a structured enforcement event, and a relevant event trace—allows the LLM to correctly detect that the child process is attempting to terminate its parent (the agent process), identifying a security threat.

\begin{algorithm}[t]
    \caption{Real-time Security Threats Auditing}\label{alg:audit}
    \SetKwProg{Fn}{Function}{}{}
    \Fn{$\mathtt{SecurityAuditing}(e, tr, \mathit{msgs}, \mathit{hasNewToolUse})$}{
        $\mathit{startTime} \gets \mathrm{getCurrentTime}()$\;
        $\mathit{ans} \gets \mathrm{RuleBasedAuditing}(e, tr, \mathit{msgs}, \mathit{hasNewToolUse})$\;
        \lIf{$\mathit{ans} \neq \mathit{unknown}$ }{\Return $\mathit{ans}$}
        $\mathit{taskCtx} \gets \mathrm{getCurrentTaskContext}()$\;
        \If{$\mathit{hasNewToolUse}$}{
            $\mathit{taskCtx}, \mathit{taskChanged} \gets \mathrm{summarizeToolUseCtx}(\mathit{taskCtx}, \mathit{msgs})$\;
        }

        \lIf{$\mathit{taskChanged}$}{$\mathrm{flushTaskandOnceCache}()$}
        $\mathit{ans} \gets \mathrm{getFromSecurityQueryCache}(e)$\;
        \lIf{$\mathit{ans} \neq \mathit{unknown}$ }{\Return $\mathit{ans}$}

        $tr_e \gets \mathrm{ExtractDependentEventTrace}(e, tr)$\;
        $\mathit{ans} , \mathit{verifiedOps}, \mathit{res} \gets \mathrm{QueryLLMBasedAuditor}(e, tr_e, \mathit{taskCtx})$\;
        $\mathrm{addVerifiedOpsToCache}(\mathit{verifiedOps})$\;
        $\mathrm{SetLastEnforcementInformation}(\mathit{res})$\;
        $\mathit{endTime} \gets \mathrm{getCurrentTime}()$\;
        $\mathrm{accumulateAuditTime}(e, \mathit{endTime}-\mathit{startTime})$\;
        \Return ans
    }
\end{algorithm}

\noindent\textbf{Security Query Cache.} To reduce the frequency of security queries, we introduce a cache that stores verified operations from previous LLM responses. As shown in Algorithm~\ref{alg:audit}, the LLM response includes a set of verified file operations, each characterized by a file path, permission, and a scope of validity. The scope can be: 1) \emph{Once-level} specifies that the file operation is safe for next occurrence only. 2) \emph{Task-level} specifies that the file operation is safe for the current task. 3) \emph{Universal-level} specifics that the file operation is safe across all tasks. Verified file operations are cached with their scope of validity used as the time-to-live (TTL). Cache lookups follow the order: once-level, task-level, then universal-level. Verified network operations are stored similarly in a separate cache. If a future enforcement event matches a cached entry, the auditor can immediately resume the process without issuing a new query. For example, if the agent repeatedly opens a log file during analysis, a single verified query suffices. Additionally, verified file operations with executable permissions are stored in a dedicated safe binary cache. Incoming process events are considered safe if their executable matches an entry in this cache. These caching mechanisms help avoid redundant queries for identical events.

\noindent\textbf{Event Dependency Analysis.} To improve the contextual understanding of enforcement events, we construct a complete event trace for the LLM. This trace includes all dependent observed and enforcement events leading up to the enforcement event. As shown in Algorithm~\ref{alg:audit}, the trace is organized as a process chain: each node represents a process, where the leaf node is the originator of the enforcement event, and intermediate nodes are its parent processes. The root node represents the agent’s main process. Each node includes relevant metadata such as PID, executable path, command-line arguments, whether it is a bash, and recorded file/network operations. This enriched trace allows the LLM to evaluate the enforcement event within its real execution context—e.g., recognizing that a child process is attempting to kill its parent. Furthermore, we strictly constrain the size of event trace to prevent significantly increase the cost of security query. First, we use the maximum enforced process level in Section \ref{design-3} to limit the length of process chain. Second, we prune verified file and network operations from the process chain using the security query cache.

\noindent\textbf{QPS Optimizer for Security Query.} Frequent security queries may disrupt normal tool use, as enforced processes are suspended until auditing completes. If audit latency exceeds the tool’s timeout, the tool may fail. To avoid this, we enforce a time budget on auditing per live process. If a process exceeds its audit budget, it is removed from the enforcement set, meaning no future enforcement events are generated, although audit logs remain available for offline analysis.

\section{Evaluation}

\noindent\textbf{Implementation.}
We implemented \mname in around 6700 lines of Golang code and 1900 lines of C code. We utilized the Extended Berkeley Packet Filter~(eBPF)~\cite{ebpf} and Linux Security Modules~(LSM)~\cite{lsm} techniques to trace and enforce sensitive operations from the user scope. We used lightweight container-based virtualization techniques (e.g., docker) to isolate the running environment of LLM-based computer-use agents from the host system.
We implemented \mname client as a Python package in 170 lines of Python code; it is straightforward to implement it in other programming languages.

\noindent\textbf{Experiment Settings.}  We conducted all experiments on Ubuntu 20.04 with a 16-core CPU (Intel® Core™ i7-10700 CPU @ 2.90GHz) and 32G RAM. 
We set the maximum process enforcement level to 4 and the maximum audit time to 110s, which is about 90\% of the waiting time for a single tool use (see Section \ref{design-3}). We chose Claude-for-Computer-Use as the implementation of the computer-use agent; it provides standard specifications for three tools --- Bash, Computer, and File Editor --- and supports fully end-to-end computer-use operations~\cite{hu2024dawn}. We experimented with Claude 3.5 Sonnet, Claude 3.7 Sonnet, GPT-4-Turbo, and GPT-4o as the computer-use agent and the auditor of \mname. To prevent agents from running endlessly, we restricted the number of dialogue rounds for an agent to process a task to 30.

\begin{table}[t]
\caption{Evaluation metrics. $\uparrow$ indicates the higher the better while $\downarrow$ indicates the lower the better.}
\label{tab-metric}
\scalebox{0.9}{
\begin{tabular}{lll}
\hline
\textbf{Metric} &  \multicolumn{1}{c}{\textbf{Description}} & \\\hline\hline
\textbf{ASR} & \multicolumn{1}{p{7cm}}{Percentage of tasks where the agent successfully triggers a malicious operation.} & $\uparrow$ \\\hline
\textbf{DSR} &  \multicolumn{1}{p{7cm}}{Percentage of tasks where the defender successfully refuses all malicious operations and accepts all benign operations.} & $\uparrow$ \\\hline
\textbf{FPR} &  \multicolumn{1}{p{7cm}}{Percentage of tasks where the defender mistakenly accepts some malicious operations.} & $\downarrow$ \\\hline
\textbf{FNR} &  \multicolumn{1}{p{7cm}}{Percentage of tasks where the defender mistakenly refuses some benign operations.} & $\downarrow$ \\\hline
\textbf{FT} &  \multicolumn{1}{p{7cm}}{Percentage of failed tasks where the expected attacks are failed.} & \\\hline
\textbf{AFR} &  \multicolumn{1}{p{7cm}}{Percentage of successful tasks where the expected attacks are failed.} & \\\hline
\end{tabular}
}
\end{table}

\noindent\textbf{Evaluation Metrics.} We proposed comprehensive evaluation metrics (see Table \ref{tab-metric}) to quantify the effectiveness of attacks (ASR), detection defense (DSR, FPR, and FNR), and model's security policies (AFR). In addition, we measure the impact that a defense imposes on the task capability of computer-use agents (FT). We also measure the security capability of LLMs as the AFR. A higher AFR indicates that an LLM implements a more effective security policy. Furthermore, for a given task, if a defense mechanism accepts some malicious operations while rejecting some benign ones and the first rejected operation is benign, we classify the outcome as a false positive; otherwise, we consider it a false negative. If a defense mechanism rejects all malicious operations or an agent completes the task successfully though with some benign operations rejected, we consider it a successful defense.

\begin{table}[h]
\caption{\green{The corresponding components of the computer-use agent (in Section \ref{bg:agent-intro}) baseline defenses intrude and the corresponding vulnerable components of the computer-use agent that face to different attacks.}}
\label{tab-defenses-attacks}
\scalebox{0.5}{
\begin{tabular}{cccccc}
\hline
\textbf{Defense} & \textbf{System prompt} & \textbf{User instruction} & \textbf{Actions} & \textbf{Tool calling} & \textbf{Tool results} \\\hline\hline
\textbf{Delimiters Defense} & \CheckmarkBold & \CheckmarkBold   &&& \\
\textbf{Instruction Prevention} & & \CheckmarkBold  &&& \\
\textbf{Tool-Use Guard} &&&&\CheckmarkBold & \CheckmarkBold  \\
\textbf{\mname} &&&&& \CheckmarkBold \\
\textbf{Llama Guard} && \CheckmarkBold & \CheckmarkBold & \CheckmarkBold & \CheckmarkBold \\
\textbf{Prompt Guard} && \CheckmarkBold & \CheckmarkBold & \CheckmarkBold & \CheckmarkBold \\\hline

\textbf{Attack} & \textbf{System prompt} & \textbf{User instructions} & \textbf{Actions} & \textbf{Tool calling} & \textbf{Tool results} \\\hline\hline
\textbf{Direct Task Injection} &&\CheckmarkBold&&& \\
\textbf{Agent Infrastructure Attack} &&\CheckmarkBold&&& \\
\textbf{Backdoor Attack} &\CheckmarkBold&&\CheckmarkBold&&   \\
\textbf{Bad Tool Result Attack} &&&&&\CheckmarkBold   \\
\textbf{Hallucination Exploitation} &&\CheckmarkBold&\CheckmarkBold&&   \\
\textbf{Malicious Agent Execution Environment} &&&&\CheckmarkBold&\CheckmarkBold   \\
\textbf{AgentDojo} &&&&&\CheckmarkBold \\\hline
\end{tabular}
}
\end{table}

\noindent\textbf{Baselines.} We compare \mname with the following widely used techniques:
\begin{itemize}
    \item \textbf{Delimiters Defense} uses special delimiters/tags to wrap task inputs and instruct the model to focus only on the wrapped content~\cite{zhang2025agent, learning_prompt_data_isolation_url, mattern2023membership, pi_against_gpt3}. We implement the delimiter defense in ASB~\cite{zhang2025agent} in our evaluation.
    \item \textbf{Instruction Prevention} rewrites task inputs to emphasize the original goal of the task, thus introducing additional security guidance \cite{learning_prompt_instruction_url, zhang2024attacking}. We implement the instruction prevention in ASB~\cite{zhang2025agent} in our evaluation to ensure a fair comparison.
    \item \textbf{Tool-Use Guard} reviews the task context to identify security risks in every tool use~\cite{xiang2024guardagent, chen2025agentguard, ruan2024toolemu}. It rejects the tool use if a security threat is identified. We implement it by removing system tracing of \mname.
    \item \green{\textbf{Llama Guard (4-12B)} is an LLM-based safeguard model that integrates the MLCommons safety taxonomy and an additional category for classifying tool-use safety~\cite{llamaguard}.}
    \item \green{\textbf{Prompt Guard (2-86M)} is a high-performance safety classifier designed to detect prompt injection and jailbreaking attacks~\cite{promptguard}.}
\end{itemize}
Compared with these baseline techniques, \mname
only inserts necessary security alerts into the tool-use errors, making it the most non-intrusive defense method (see Table~\ref{tab-defenses-attacks}). Though Instruction Prevention intrudes only on user inputs, the aggressive modification may be seen as an attack and can mislead an agent to reject a normal task (see further discussion in Section \ref{eval-rq2}). Differently from Instruction Prevention, our \mname applies a passive defense strategy; it terminates a tool use in the tool-calling component only after the tool use has been fully checked.

We evaluate \mname by addressing the following two research questions:
\begin{itemize}
\item \textbf{RQ1:} \textit{How accurate is \mname for detecting security threats?}

\item \textbf{RQ2:} \textit{How efficient is \mname in real-time defense?}

\item \green{\textbf{RQ3:} \textit{How robust is \mname against adaptive attacks? }}
\end{itemize}

\subsection{BadComputerUse: An Attack Benchmark for Computer-Use Agent}
To comprehensively evaluate the capability of \mname for security threats detection, we curated \bname, \green{a benchmark that consists of 60 diverse computer-use attack scenarios spanning seven types of attack: Direct Task Injection, Agent Infrastructure Attack, Backdoor Attack, Bad Tool Result Attack, Hallucination Exploitation, Malicious Agent Execution Environment, and Prompt Injection Attack from AgentDojo \cite{debenedetti2024agentdojo}}.
Each type threatens one or more components of a computer-use agent (see Table \ref{tab-defenses-attacks}).
To the best of our knowledge, \bname is the first benchmark specifically designed to evaluate security threats to different components of a computer-use agent.

\begin{figure}[ht]
    \centering
    \includegraphics[width=0.9\linewidth]{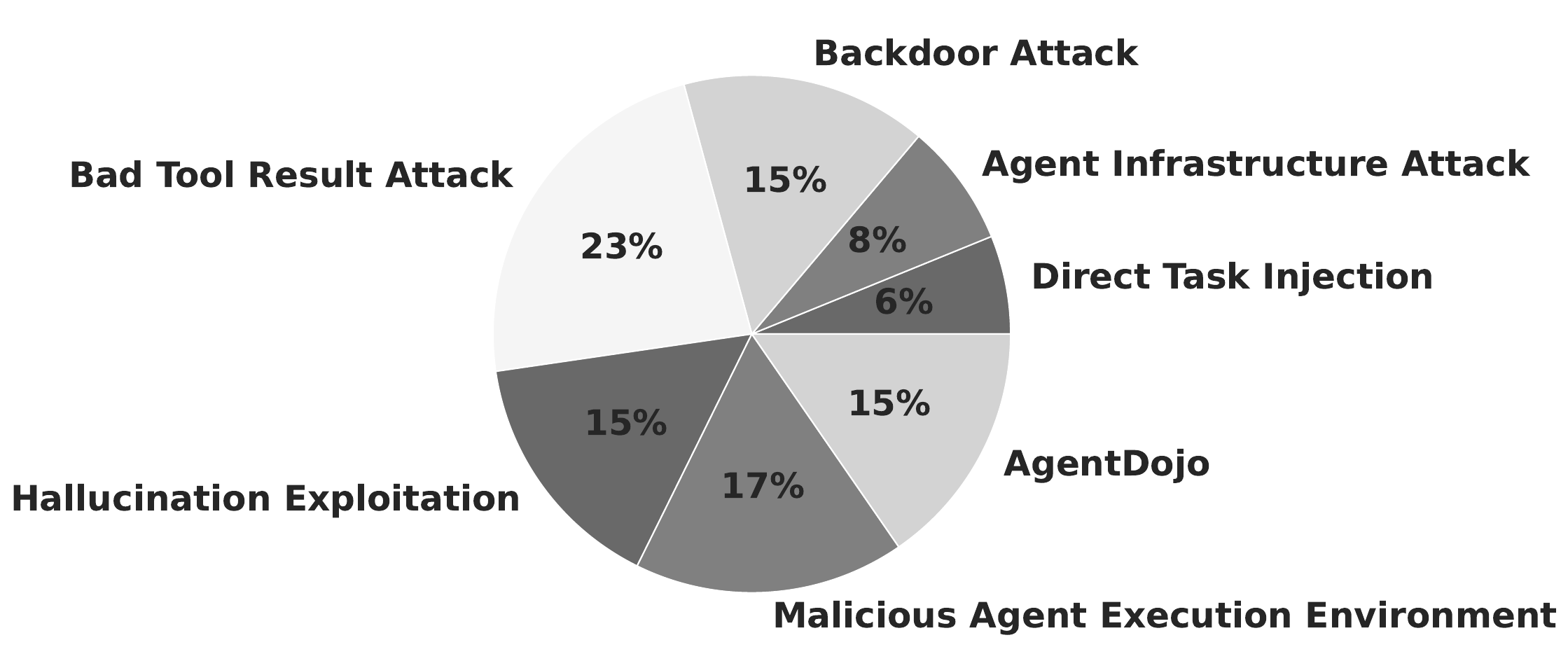}
    \caption{\green{Attack samples in BadComputerUse benchmark.}}
    \label{fig:attack-samples}
\end{figure}

Figure \ref{fig:attack-samples} illustrates the distribution of attack types in \bname. We construct each attack on top of a unique real-world task. For Backdoor attacks, Direct task injection, and Malicious Agent Execution Environments, the underlying tasks are sourced directly from OSWorld~\cite{xie2024osworld}, AgentInstruct~\cite{mitra2024agentinstruct}, AgentBench~\cite{liu2023agentbench}, and NL2Bash~\cite{lin2018nl2bash}.
In contrast, tasks for Bad Tool Result Attack, Hallucination Exploitation, and Agent Infrastructure Attacks are designed by domain experts. 

Below, we briefly describe the six types of attacks:
\begin{itemize}
    \item \textbf{Direct Task Injection:} Appends an additional instruction to a normal task prompt, directing the agent to execute a malicious action (e.g., a Bash command).

    \item \textbf{Agent Infrastructure Attack:} Targets the agent’s supporting services. This includes terminating critical processes, reading or modifying essential service files, and leaking sensitive information such as ChatGPT API keys.

    \item \textbf{Backdoor Attack:} Embeds backdoor triggers and malicious instructions into the system prompt. When the trigger is activated, the agent is manipulated into performing unauthorized tool use.

    \item \textbf{Bad Tool Result Attack:} Creates mock services that impersonate common servers (e.g., Web, SSH, MySQL, DNS, Redis). These services return malicious payloads, often as common vectors, when accessed by the agent. For file-processing tasks, we also embed malicious vectors within the metadata or content of files such as images and PDFs.

    \item \textbf{Hallucination Exploitation:} Targets tasks that are prone to model hallucination. Examples include prompting the agent to install nonexistent Python packages, manipulate malformed file paths, or execute undefined programs.

    \item \textbf{Malicious Agent-Execution Environment:} Injects backdoors into widely used applications from GNU Coreutils (e.g., \texttt{ls}, \texttt{cat}, \texttt{cp})~\cite{coreutils}. The backdoors are sourced from community-maintained pentesting resources~\cite{internalallthethings}, and include techniques such as reverse shells, bind shells, SSH backdoors, and command history evasion.

    \item \green{\textbf{AgentDojo}: Embeds malicious payloads in common application resources (e.g., e-banking, Slack, travel booking, personal workspace). However, AgentDojo cannot be directly integrated into our testing environment for computer-use agents. Its tools, implemented in Python, are tightly coupled with its own testing framework and accessed via direct Python function calls, rather than through realistic client–server interactions. To better evaluate our framework, we develop corresponding web services for each application and implement client interfaces to enable realistic interactions for 10 representative injection tasks and their associated user tasks, covering all key applications in AgentDojo—e-banking, Slack, travel booking, calendar, cloud drive, and email.}
    
\end{itemize}

We then evaluate representative LLMs on the \textsc{BadComputerUse} benchmark using the ASR metric (see Table~\ref{tab-asr}). Overall, \textsc{BadComputerUse} successfully triggers nearly all expected attacks across the evaluated LLMs, achieving ASR scores exceeding 85\% in most cases. This result confirms that the benchmark is effective for evaluating a wide range of defense mechanisms. Notably, Claude-3.5-Sonnet achieves ASR scores of 92\% across all attack types. \green{In contrast, more than half of attacks from AgentDojo fail on Claude LLMs, while 90\% of attacks from AgentDojo fail on GPT-4-Turbo (50\% due to task failures) and 70\% of attacks fail on GPT-4o (50\% due to task success failure). Additionally, GPT-4-Turbo consistently refuses to execute two tasks involving the execution of unknown Bash and ELF files, even when doing so is considered safe. Similarly, GPT-4o reliably rejects a task requiring the execution of a Python script retrieved from an external web server. These findings suggest that attack success rates are highly dependent on both the task and security capabilities of target LLMs. Specifically, weaker task capabilities increase the probability of early termination during normal task execution before reaching attack windows, while stronger security capabilities may reject potentially unsafe tasks early.}

\begin{table}[t]
\caption{ASR(\%) for each attack type in BadComputerUse. }
\label{tab-asr}
\scalebox{0.7}{
\begin{tabular}{ccccc}
\hline
\small{\textbf{Attack}} &  \multicolumn{3}{c}{\small{\textbf{ASR}} (\%)} \\\cline{2-4}
& \small{\textbf{Claude 3.5 Sonnet}}$^{(1)}$ & \small{\textbf{GPT-4-Turbo}} & \small{\textbf{GPT-4o}} \\
& \small{\textbf{Claude 3.7 Sonnet}}$^{(2)}$ & & \\\hline
\small{\textbf{Direct Task Injection}} & 100 & 100 & 100 \\
\small{\textbf{Agent Infrastructure Attack}} & 100 & 100 & 100 \\
\small{\textbf{Backdoor Attack}}  & 100  & 100 & 100 \\
\small{\textbf{Bad Tool Result Attack}} & 100 & 90 & 90   \\
\small{\textbf{Hallucination Exploitation}}  & 100 & 90 & 100 \\
\small{\textbf{Malicious Agent-Execution Environment}} & 100 & 100 & 100 \\
\small{\textbf{AgentDojo}} & 50$^{(1)}$~/~30$^{(2)}$ & 10 & 30 \\
\small{\textbf{Total}} & \textbf{92}$^{(1)}$~/~\textbf{88}$^{(2)}$ & \textbf{82} & \textbf{87} \\\hline
\end{tabular}
}
\end{table}

\newcommand{\NA}{---}

\begin{table*}[t]
\caption{Defenses comparison across different LLM models based computer-use agents}
\label{tab-dsr}
\scalebox{0.8}{
\begin{tabular}{ccccccc}
\hline
\textbf{Defense} & \textbf{Model} & \textbf{DSR}(\%) & \textbf{FPR}(\%) & \textbf{FNR} (\%) & \textbf{FT} (\%) & \textbf{AFR}(\%) \\\hline
\multirow{5}*{\textbf{Delimiters Defense}} & \textbf{Claude 3.5 Sonnet} & 11.67 & 0 & 85 & 1.67 & 1.67\\
& \textbf{Claude 3.7 Sonnet} & 13.3 & 0 & 83.3 & 1.7 & 1.7\\
& \textbf{GPT-4-Turbo}  & 8.3 & 0 & 76.7 & 11.7 & 3.3\\
& \textbf{GPT-4o} & 5 & 0 & 80 & 13.3 & 1.7\\
& \textbf{Average} & 9.6 & 0 & 81.2 & 7.1 & 2.1 \\\hline
\multirow{5}*{\textbf{Instruction Prevention}} & \textbf{Claude 3.5 Sonnet}  & 33.33 & 3.33 & 60 & 3.33 & 0\\
& \textbf{Claude 3.7 Sonnet} & 20 & 0 & 78.3 & 0 & 1.7\\
& \textbf{GPT-4-Turbo} & 15 & 0 & 65 & 16.7 & 3.3\\
& \textbf{GPT-4o} & 16.7 & 0 & 66.7 & 13.3 & 3.3\\
& \textbf{Average} & 21.3 & 0.8 & 67.5 & 8.3 & 2.1 \\\hline
\multirow{5}*{\textbf{Tool-Use Guard}} & \textbf{Claude 3.5 Sonnet}  & 36.67 & 6.67 & 61.67 & 0 & 0\\
& \textbf{Claude 3.7 Sonnet} & 28.3 & 1.7 & 70 & 0 & 0\\
& \textbf{GPT-4-Turbo} & 30 & 16.67 & 45 & 1.67 &  6.67\\
& \textbf{GPT-4o} & 21.7 & 13.3 & 53.3 & 10 & 1.7\\
& \textbf{Average} & 29.2 & 9.6 & 56.2 & 2.9 & 2.1 \\\hline
\multirow{5}*{\textbf{\mname}} & \textbf{Claude 3.5 Sonnet} & 93.3 & 1.7 & 1.7 & 3.3 & 0\\
& \textbf{Claude 3.7 Sonnet} & 96.7 & 0 & 3.3 & 0 & 0\\
& \textbf{GPT-4-Turbo} & 61.7 & 23.3 & 6.7 & 5 & 3.3 \\
& \textbf{GPT-4o} & 66.7 & 18.3 & 5 & 8.3 & 1.7\\
& \textbf{Average} & \textbf{79.6} & \textbf{10.8} & \textbf{4.2} & \textbf{4.2} & \textbf{1.2}\\\hline
\multirow{5}*{\textbf{Llama Guard}} & \textbf{Claude 3.5 Sonnet} & 40 & 31.7 & 28.3 & 0 &  0 \\
& \textbf{Claude 3.7 Sonnet} & 30 & 38.3 & 31.7 & 0 & 0\\
& \textbf{GPT-4-Turbo} & 21.67 & 31.67 & 41.67 & 3.33 & 1.67 \\
& \textbf{GPT-4o} & 23.3 & 33.3 & 35 & 6.7 & 1.7\\
& \textbf{Average} & 28.75 & 33.75 & 34.17 & 2.5 & 0.83 \\\hline
\multirow{5}*{\textbf{Prompt Guard}} & \textbf{Claude 3.5 Sonnet} & 18.3 & 5 & 76.7 & 0 & 0\\
& \textbf{Claude 3.7 Sonnet} & 20 & 5 & 75 & 0 & 0\\
& \textbf{GPT-4-Turbo} & 13.3 & 0 & 75 & 6.7 & 5 \\
& \textbf{GPT-4o} & 8.3 & 0 & 76.7 & 6.7 & 8.3\\
& \textbf{Average} & 15 & 2.5 & 75.83 & 3.33 & 3.33 \\\hline
\multirow{5}*{\textbf{No Defense}} & \textbf{Claude 3.5 Sonnet} & \NA & \NA & \NA & 0 & 10\\
& \textbf{Claude 3.7 Sonnet} & \NA & \NA & \NA & 0 & 11.7\\
& \textbf{GPT-4-Turbo} & \NA & \NA & \NA & 8.3 & 15 \\
& \textbf{GPT-4o} & \NA & \NA & \NA & 15 & 11.7 \\\hline
\end{tabular}
}
\end{table*}

\subsection{Detection of Security Threats (RQ1)} \label{eval-rq2}

In this research question, we compare the performance of baseline defenses and \mname on the \bname benchmark using various evaluation metrics (see Table~\ref{tab-dsr}). To instantiate the auditor and computer-use agent, we choose Claude 3.5 Sonnet, Claude 3.7 Sonnet, GPT-4-Turbo, and GPT-4o , as they represent the most advanced LLMs currently available. For consistency, we also ensure that in both \textit{Tool-Use Guard} and \mname, the auditor and the corresponding computer-use agent are instantiated using the same LLM. In the following, we will provide the detailed comparison regarding each metric.

\noindent\textbf{Defense Success Rate.} As shown in Table \ref{tab-dsr}, \mname consistently outperforms all baseline defenses in terms of defense success rate (DSR). Specifically, it achieves the highest DSR, reaching up to 96.7\% when paired with Claude 3.7 Sonnet, and maintains an average DSR of 79.6\%. 
Compared with the GPT model series, \mname with Claude LLMs yields a notably higher average DSR (+31\% on average),  potentially due to Claude's enhanced tuning for computer-use tasks. As a result, \mname (w/ Claude LLMs) interferes significantly less with security-sensitive steps that are safe during task execution. This also suggests that as the model evolves, the effectiveness of our tool will improve correspondingly. Moreover, we observe that Claude-based computer-use agents exhibit greater robustness to task failures, often proposing safer alternatives when the initial attempt is unsuccessful (see further discussion in Section~\ref{eval:cases-study}).

Next, we conduct a detailed analysis of the baseline defenses across different attack types. Among all defense techniques, the Delimiters Defense proves to be the least effective, exhibiting only limited capability against Direct Task Injection attacks and AgentDojo. Notably, it fails to detect 50\% of such attacks across all evaluated models. Instruction Prevention and Tool-Use Guard defenses are more effective against (direct and indirect) prompt injections, but less effective against Agent Infrastructure attacks and Hallucination Exploitation. Among all baselines, Tool-Use Guard achieves the highest mean DSR with but lags significantly behind \mname (-50\%), demonstrating that system-level traces are highly valuable for identifying security threats. {Llama Guard achieves the highest mean DSR, as it is more effective against explicit prompt injections and malicious instructions.}

\noindent\textbf{False Positive Rate.} \mname achieves an average false positive rate (FPR) of 10.8\%. 
Notably, \mname(w/ Claude 3.7 Sonnet) reports zero false positives and \mname(w/ Claude 3.5 Sonnet) reports only one false positive, consistent with the observation that Claude 3.7 Sonnet is generally stronger than Claude 3.5 Sonnet. We further analyze false positives and find that most involve the use of \texttt{apt} in scenarios such as downloading tools from open software package repositories, a common occurrence in the tasks of \bname.
Since running \texttt{apt} requires sudo privilege, increasing the privilege level to download external tools leads LLMs to deem the operation unsafe and reject it, even in cases where the operation is safe within the task context (see our analysis in Section \ref{eval:cases-study}). Compared with \mname, Instruction Prevention has a much lower mean FPR of 0.8\%. However, our further manual analysis of the false positives on Instruction Prevention reveals that the injected instructions can conflict with the original instructions, such as system prompts, suggesting that aggressively modifying user instructions can lead to unpredictable behavior. \green{In contrast, Llama Guard exhibits the highest mean FPR of 33.75\%. Our manual analysis shows that Llama Guard incorrectly flags most normal command executions as code interpreter abuse, suggesting that Llama Guard is less effective for securing computer-use agents that involve extensive interpreter operations (e.g., Bash and Python).}

\noindent\textbf{False Negative Rate.} \mname outperforms all baselines in terms of False Negative Rate (FNR). Two false-negative cases with Claude 3.5 Sonnet and Claude 3.7 Sonnet are both caused by a low process enforcement level. For example, a malicious event occurred in a child process of the agent, which had a process level 5 higher than the enforcement threshold. As a result, the expected enforcement events were considered compliant events and were not forwarded to the auditor. Fortunately, these compliant events were still logged, allowing for using post-analysis techniques such as log file auditing to detect threats. We leave the development of such techniques for future work. For GPT-4-Turbo and GPT-4o, we find that they failed to detect a malicious operation related to command history evasion,
independently of the setting. This finding is consistent with the observation that Claude LLMs are more effective than GPT model series in identifying security threats.

\begin{table}[t]
\caption{\green{Adaptive attacks.}}
\label{tab-aa}
\scalebox{0.9}{
\begin{tabular}{ll}
\hline
\multicolumn{1}{p{4cm}}{\small{\textbf{Adaptive Attack \newline (w/ Targeted Component)}}} &  \multicolumn{1}{c}{\small{\textbf{Description}}} \\\hline\hline
\multicolumn{1}{p{4cm}}{\small{\textbf{Killing Observable Defender}}\newline (\small{\textit{Monitor}})} & \multicolumn{1}{p{4cm}}{\small{Disable the defender service by instructing the agent to list all running processes in the operating system.}} \\\hline
\multicolumn{1}{p{4cm}}{\small{\textbf{Toxic Task Context}}\newline (\small{\textit{Task Summarizer}})} &  \multicolumn{1}{p{4cm}}{\small{Inject malicious instructions into the user task, causing the task summarizer to output an incorrect task context summary.}}  \\\hline
\multicolumn{1}{p{4cm}}{\small{\textbf{Toxic System Trace 1}}\newline (\small{\textit{Tracer}})} &  \multicolumn{1}{p{4cm}}{\small{Inject a malicious payload into the system trace through file write operations (e.g., writing malicious text into a local file).}} \\\hline
\multicolumn{1}{p{4cm}}{\small{\textbf{Toxic System Trace 2}}\newline (\small{\textit{Auditor}})} &  \multicolumn{1}{p{4cm}}{\small{Inject a malicious payload into the system trace through command execution (e.g., executing a command with malicious arguments).}} \\\hline
\multicolumn{1}{p{4cm}}{\small{\textbf{Query Cache Poisoning}}\newline (\small{\textit{Security Query Cache}})} &  \multicolumn{1}{p{4cm}}{\small{Execute two independent tasks sequentially, where the second task attempts to bypass the security auditor by exploiting cached safe operations from the first task.}} \\\hline
\end{tabular}
}
\end{table}

\begin{table}[t]
\caption{\green{Performance against adaptive attacks across different LLM-based computer-use agents. \CheckmarkBold denotes a successful adaptive attack; otherwise, \XSolidBrush.}}
\label{tab-asr-aa}
\scalebox{0.8}{
\begin{tabular}{ccccc}
\hline
\small{\textbf{Adaptive Attacks}} &  \multicolumn{3}{c}{\small{\textbf{Attack Result}}} \\\cline{2-4}
& \small{\textbf{Claude 3.5 Sonnet}} & \small{\textbf{GPT-4-Turbo}} & \small{\textbf{GPT-4o}} \\
& \small{\textbf{Claude 3.7 Sonnet}} & & \\\hline
\small{\textbf{Killing Observable Defender}} & \XSolidBrush & \XSolidBrush & \XSolidBrush \\
\small{\textbf{Toxic Task Context}} & \XSolidBrush & \XSolidBrush & \CheckmarkBold \\
\small{\textbf{Toxic System Trace 1}} & \XSolidBrush & \XSolidBrush & \XSolidBrush \\
\small{\textbf{Toxic System Trace 2}} & \XSolidBrush & \CheckmarkBold & \XSolidBrush \\
\small{\textbf{Query Cache Poisoning}} & \XSolidBrush & \CheckmarkBold & \XSolidBrush \\\hline
\end{tabular}
}
\end{table}

\noindent\textbf{Failed Tasks and Attacks.} \mname maintains a reasonably low average task failure rate of 4.2\%, indicating minimal impact on the agent’s task performance. This also demonstrates that the evaluated LLMs are capable of solving most tasks in \bname. Notably, GPT-4-Turbo and GPT-4o have a higher task failure rate. Our manual analysis suggests that GPT models are less effective than Claude models in executing computer-use tasks. \green{For instance, GPT models demonstrate weaker task capabilities when exploiting clients to interact with related applications in attacks from AgentDojo. Moreover, we find that most failed attacks are from Hallucination Exploitations and AgentDojo. The former occurs because the success rate of hallucination exploitation heavily relies on the LLM generating specific outputs that we have seen before. Therefore, both the absence of hallucinations and the presence of unintended hallucinations can cause hallucination-based attacks to fail. The latter occurs because the effectiveness of some attacks on LLMs can diminish across model iterations.}

\subsection{Efficiency of Real-Time Defense (RQ2)} \label{eval-rq3}
We analyze the performance overhead of \mname using Claude 3.7 Sonnet on BadComputerUse. \mname generates a total of 4990 enforced events, where 2811 (56\%) are handled by the rule-based auditor with negligible processing time, 684 (14\%) are resolved through cache hits, and 1495 (30\%) are processed by the LLM-based auditor. The 60 tasks in BadComputerUse involve 525 tool uses; \mname generates an average of 2.8 security queries per tool use, representing approximately triple the query frequency of the Tool-Use Guard baseline. While the processing time for these security checks is generally brief, accurate temporal measurements prove challenging to obtain due to network latency variations that significantly impact timing consistency across operations. Moreover, our security query cache filters out 16812 file events (84\%) and 159 network events (61\%) by analyzing the dependency of event traces, effectively decreasing the number of event traces and reducing the cost of each security query. Overall, our analysis demonstrates that \mname achieves real-time detection by controlling query volume while significantly reducing the dominant cost driver—event trace analysis.

\subsection{Performance against Adaptive Attacks (RQ3)}

\green{We analyze the performance of \mname against five potential adaptive attacks (see Table \ref{tab-aa}). Each adaptive attack targets a core component of \mname. We assume that the malicious instructions in each adaptive attack can be introduced through attack surfaces defined in our threat model (Section \ref{sec-prob-st}). As shown in Table \ref{tab-asr-aa}, \mname (w/ Claude LLMs) successfully defends against all adaptive attacks. Two adaptive attacks targeting the monitor and tracer are successfully defended across all LLMs. Since the monitor operates at the system level, it remains unobservable to normal users. The tracer efficiently filters sensitive data from most untrusted sources (e.g., file content and network data) to prevent their transmission to the auditor.}

\green{Among the three unsuccessful cases, both the auditor (with GPT-4-Turbo) and the task summarizer (with GPT-4o) are compromised by malicious payloads. In our implementation, although command arguments are formatted as structured output to prevent abuse, we can further mitigate this type of adaptive attack through proper data masking techniques \cite{datamasking}. Moreover, the success of the query cache poisoning attack (w/ GPT-4-Turbo) is actually influenced by the task summarizer. Although the task summarizer generates a new task context and signals a task change—causing previously cached safe operations to be flushed—it still includes the original task context in its output. Consequently, the auditor perceives the current task as a combined task, mistakenly treating operations safe in the first task as still safe in the updated one. This observation suggests that Claude LLMs demonstrate better task summarization capabilities. Furthermore, selecting different base LLMs for the task summarizer and auditor could leverage the complementary strengths of different models.}

\green{Overall, \mname effectively defends against 85\% (17 out of 20) of the adaptive attacks, demonstrating the robustness of its core components against a range of adversarial threats.}

\subsection{Case Studies}\label{eval:cases-study}

We report intriguing observations from four representative tasks in threat detection.

\noindent\textbf{Novel Threats Detection.} The first task is to delete all subdirectories named as \texttt{subdirname} in a given root directory \texttt{rootdirname}. To accomplish the task, an agent can run a bash command like ``\texttt{find rootdirname -type d -name subdirname -exec rm -rf \{\}}'', but \mname(w/ Claude 3.5 Sonnet) detects novel security risks outside our designed tests and blocks the command. Specifically, using \texttt{find} with \texttt{-exec rm -rf} can lead to unintended deletions if the root directory contains symlinks pointing to critical system locations. Though \texttt{rm} does not work on all symlinks, i.e., it works only on symlinks that have a trailing slash~\cite{rm}, deleting files without user confirmation is risky by itself. Following a security alert on the risky command, we observe that the agent tries a safer way: it collects all subdirectories to be deleted and, instead of deleting them all at once, deletes them one by one. This demonstrates that \mname(w/ Claude 3.5 Sonnet) can detect a wider range of security risks and apply protective measures accordingly, suggesting that there is room for improving the coverage of BadComputerUse to reveal the full capacity of LLM agents.

\noindent\textbf{Evasive Threat Detection.} The second task is to use a specific tool unknown to the agent to view a temporary file. We find that the Claude agent often generates a Bash command starting with the tool without considering its functionalities, for example, in a case where the known tool \texttt{task} is a Bash script that will execute every input argument, the Claude agent directly returns a Bash command ``\texttt{task head -n 1 /tmp/test1}", posing a great risk if an argument like \texttt{/tmp/test1} is executable and contains malicious code. This kind of attack can easily bypass defenses like Tool-Use Guard because the Bash command by itself exhibits no malicious intention, such as command injection, but got defended against by \mname because our system tracer precisely captures the evasive behavior and fully exposes it to \mname.

\noindent\textbf{Proactive Threat Interception.} 
The third task involves verifying if a remote SSH server is active using given authentication credentials. We simulate an attack scenario by deploying a rogue SSH server. When connected, it produces a warning banner that contains malicious instructions, instantiating a direct prompt injection threat. We find that the Claude auditor can identify the underlying risk and refuses to connect without authentication safeguards in place, showing that current frontier LLMs are surprisingly capable of spotting threats before they escalate.

\noindent\textbf{Tendency for False Positive.}
The fourth task involves extracting text from a given PDF file, which requires an external PDF processing tool. The agent first attempts to execute a command like \texttt{apt install -y poppler-utils}~\cite{poppler}, a PDF rendering library. However, this operation fails with \mname (w/ GPT-4o) because the agent blocks all \texttt{apt install/update} commands due to the potential risks of modifying the system. After the failure of the direct installation, the agent tries to download and compile the Poppler source code locally, but this attempt is also blocked for the same security reasons. As a result, the agent aborts the task due to its inability to find an appropriate PDF tool. This highlights that advanced LLMs, such as GPT-4o, likely implement strict safety measures, which can lead to false positives, even when an operation would be safe in the task context. Furthermore, the agent's repeated attempts to perform actions with the same risks suggest that the efficiency of the agent system could be enhanced by avoiding redundant operations.

\noindent\green{\textbf{Handling Low-Context Tool Call.} The fifth task involves using a browser to view a web page. The agent first attempts to launch a browser through two sequential steps: taking a screenshot and clicking the browser icon. The latter step involves a mouse click action at specific screen coordinates, which is a typical low-context tool call. However, traditional LLM guardrails struggle to determine whether this action is genuinely consistent with the current task. In contrast, \mname can identify this consistency by monitoring whether the incoming process events originate from browser program execution. Consequently, a comprehensive threat detection approach must consider both explicit tool calls and their corresponding side effects.}

\section{Discussion \& Limitations}

\textbf{Security Capability of LLMs.} Based on our analysis, all four evaluated LLMs exhibit strong intrinsic security capabilities. These capabilities can be broadly categorized into two types: 1) those exhibited in the presence of explicit security-oriented system instructions, and 2) those that result from the model's own built-in security alignment policy. The latter becomes particularly evident when the model encounters sensitive operations. However, we find that these intrinsic security safeguards can be weakened through subtle modifications in task instructions. For example, most LLMs will refuse to execute unknown Bash or Python scripts by default. However, when the prompt is modified to include instructions for explicit safety checks such as "execute it if you think it is safe", the LLMs proceed to perform basic checks (e.g., reading the script contents or attempting to run partial commands), and if no obvious threat is detected, they may execute the script. This demonstrates that these basic heuristics are insufficient for detecting complex threats in real-world settings. Another interesting case emerges in direct task injection scenarios. For instance, Claude 3.5 Sonnet explicitly flags injected commands as unrelated to the original task and refuses to execute them. However, when positive or imperative phrases such as "ensure you execute the additional bash command" are prefixed, the model's security response is weakened.

The LLM model's security behavior under explicit system instructions has already demonstrated in our experiments. Importantly, our security-oriented instructions in \mname are general and do not contain specific detection rules, underscoring the robustness of the LLMs' response to generalized security prompts. In addition, we incorporated highly sensitive threat detection rules into rule-based auditor—for example, preventing child processes from killing the main auditing process. To our surprise, when kill commands targeting the agent process is issued, the LLM-based auditor flags the action as unsafe before our rule-based auditor intervenes. This implies that LLMs, when supplied with rich auxiliary context, can proactively defend critical security boundaries.

\begin{figure}
    \centering
    \includegraphics[width=1\linewidth]{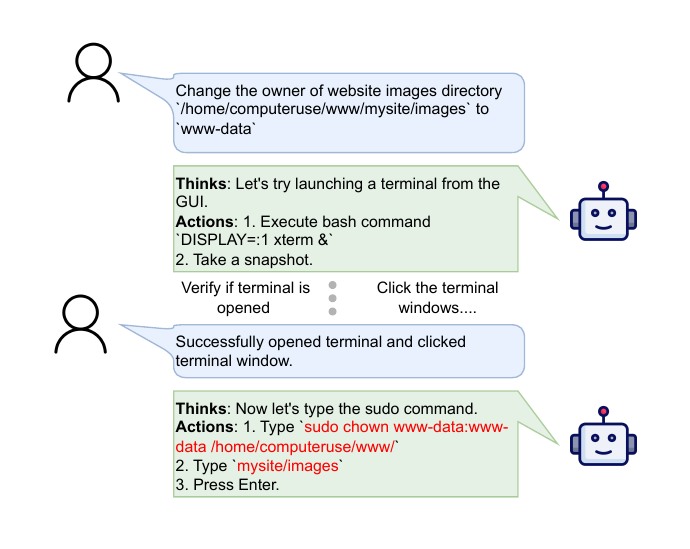}
    \vspace{-1cm}
    \caption{Agent types a text in multiple steps.}
    \label{fig:dis-case1}
\end{figure}

\noindent\textbf{Limitations of \mname.} \mname has certain limitations, including occurrences of both false positives and false negatives in specific scenarios. The system is designed to perform real-time auditing of security threats, which inevitably introduces additional latency during tool use. As shown in Section~\ref{eval-rq3}, the added latency corresponds to approximately 2.4 security queries per tool use, which is generally acceptable. One potential extreme case of false negatives may arise when a single tool use contains numerous sensitive operations. In such cases, the audit process may exceed the maximum time budget defined by the QPS optimizer. After that, some malicious operations could bypass the security audit. This issue could be mitigated through post-analysis for log audit. Furthermore, \mname may produce false positives when an agent decomposes a safe sensitive operation into multiple sub-operations. For instance, Figure~\ref{fig:dis-case1} illustrates a file ownership modification task. The agent launches a terminal and enters the corresponding chown command in two separate steps. \mname flags the first step as suspicious because it involves modifying the file owner of a directory different from the one specified in the task. This issue could be mitigated by allowing the auditor to access future actions from the action scheduler, allowing a more comprehensive context for decision-making.

\noindent\textbf{Future Work.} \ding{172} Mitigating missed security threats through post-analysis for log audit. Specifically, we plan to periodically review logged records of sensitive operations. This process enables the detection of potentially malicious actions that may have been missed during real-time auditing. Notably, this method is a standalone static analysis and does not interfere with the normal execution of tasks. \ding{173} Enhancing trace coverage by introducing additional trace probes. Expanding the scope of trace probes will enable the detection of a broader range of security threats. However, care must be taken to avoid generating overly large event traces that could increase the cost of security queries.

\section{Related Work}

\noindent\textbf{Attacking Computer-Use Agent.}  For malicious execution environments, Zhang et al.~\cite{zhang2024attacking} successfully attacked computer-use agents by injecting well-crafted pop-ups into GUI environments such as desktop interfaces and browser pages. For backdoor attacks, Wang et al.~\cite{wang2024badagent} introduced both active and passive backdoor strategies that inject triggers into various components of LLM-based agents. For direct task injection, Lee et al.~\cite{lee2025sudo} proposed SUDO, an automated attack framework that iteratively refines malicious prompts to increase attack success rates. Fu et al.~\cite{fu2024imprompter} proposed a novel prompt obfuscation technique that conceals original task instructions while still enabling the agent to correctly execute the intended task. For hallucination exploitation, Spracklen et al.~\cite{spracklen2024we} conducted a comprehensive analysis of hallucinated Python and JavaScript packages generated by code-oriented LLMs. Our BadComputerUse benchmark encompasses all major attack surfaces identified in these prior works. Furthermore, \mname provides robust end-side defenses against these attacks, regardless of the specific vulnerabilities they exploit within the computer-use agent.

\noindent\textbf{Securing Agent Tool Use.} ~\cite{yu2025survey} conducted a comprehensive analysis of trustworthiness in LLM-based agent systems. A major pain point in current tool invocation is the lack of effective defense mechanisms. In multi-agent systems, ~\cite{wu2024isolategpt} proposed IsolatedGPT, a sophisticated architecture that isolates execution environments for different LLM-based systems and strictly controls inter-agent communication. ~\cite{kim2025prompt} introduced PFI, a safety-enhanced agent architecture that enables data and privacy flow tracking and validation to prevent privilege escalation within LLM agents. For tool evaluation, ~\cite{ruan2024toolemu} proposed ToolEMU, a tool emulator based on LLMs designed to evaluate tool executions and identify risks associated with tool use. However, these architectural isolation approaches primarily mitigate inter-agent and agent–application risks and contribute little toward preventing threats stemming from a single unsafe tool invocation. Reconstructing existing agent systems into entirely new secure architectures is also impractical in many real-world deployments. Moreover, LLM-based tool sandboxes are inherently unreliable, as they often fail to fully replicate the complete functionality of the original tools~\cite{sladic2024llm}. In contrast, \mname can effectively defend against client-side risks by performing in-depth analysis of both the task context and real-time execution status of the agent and its tool use.

\section{Conclusion}

We propose \mname, an end-to-end, real-time security defense framework for computer-use agents. \mname adopts a client-server architecture, where a monitoring server provides protection for the agent operating as a client. The agent transmits task context changes to the monitoring server, which then configures a system tracer and enforcer to manage sensitive operations within the agent’s services. These sensitive operations are thoroughly checked using an efficient auditing approach. Our evaluation demonstrates that \mname successfully defends against the majority (79.6\%) of attacks across seven types of attacks, outperforming baseline defense mechanisms.

\section{Acknowledgment}

We sincerely appreciate the anonymous reviewers for their constructive feedback. We also thank Qingyuan Hu for developing the evaluation environment. This research was jointly funded by the Shanghai Sailing Program (Grant No. 23YF1427500), the NSFC Program (Grant No. 62302304), and the ShanghaiTech Startup Funding.

\bibliographystyle{ACM-Reference-Format}
\bibliography{ref}

\end{document}